%% file: main.tex
\documentclass[11pt]{article}
\usepackage{UF_FRED_paper_style}
\usepackage{xcolor}
\usepackage{lipsum}  %% Package to create dummy text (comment or erase before start)
\usepackage{amsmath}
\usepackage{comment}

\newcommand{\be}{\begin{equation}}
\newcommand{\ee}{\end{equation}}
\newcommand{\bea}{\begin{eqnarray}}
\newcommand{\eea}{\end{eqnarray}}

\title{Toxic Liquidation Spirals
}

\author{Jakub Warmuz 
    \thanks{jakub@0vix.com}
\and Amit Chaudhary 
     \thanks{a.chaudhary.1@warwick.ac.uk} %% Email author 1 
\and Daniele Pinna
     \thanks{phys2172@ox.ac.uk / daniele@0vix.com} %% Email author 2
    }
    
\date{December 14, 2022}

\begin{document}
{\setstretch{.8}
\maketitle
% %%%%%%%%%%%%%%%%%%
\begin{abstract}
% CONTENT OF ABS HERE--------------------------------------

On November 22nd 2022, the lending platform AAVE v2 (on Ethereum) incurred bad debt resulting from a major liquidation event involving a single user who had borrowed close to \$40M of CRV tokens using USDC as collateral. This incident has prompted the Aave community to consider changes to its liquidation threshold, and limitations on the number of illiquid coins that can be borrowed on the platform. In this paper, we argue that the bad debt incurred by AAVE was not due to excess volatility in CRV/USDC price activity on that day, but rather a fundamental flaw in the liquidation logic which triggered a toxic liquidation spiral on the platform. We note that this flaw, which is shared by a number of major DeFi lending markets, can be easily overcome with simple changes to the incentives driving liquidations. We claim that halting all liquidations once a user's loan-to-value (LTV) ratio surpasses a certain threshold value can prevent future toxic liquidation spirals and offer substantial improvement in the bad debt that a lending market can expect to incur. Furthermore, we strongly argue that protocols should enact dynamic liquidation incentives and closing factor policies moving forward for optimal management of protocol risk.

% END CONTENT ABS------------------------------------------
\noindent
\noindent
\end{abstract}
}

% %%%%%%%%%%%%%%%%%%%%%%%%%%%%%%%%%%%%%%%%%%%%%%%%%%%%%%%%%%
% %%%%%%%%%%%%%%%%%%%%%%%%%%%%%%%%%%%%%%%%%%%%%%%%%%%%%%%%%%
% BODY OF THE DOCUMENT
% %%%%%%%%%%%%%%%%%%%%%%%%%%%%%%%%%%%%%%%%%%%%%%%%%%%%%%%%%%
% %%%%%%%%%%%%%%%%%%%%%%%%%%%%%%%%%%%%%%%%%%%%%%%%%%%%%%%%%%

% --------------------
\section{Introduction}
\label{sec:01}
\input{sections/01_Intro}
% --------------------
\section{Liquidation Mechanics}
\label{sec:02}
\input{sections/02_Liquidation_Mechanics}
% --------------------
\section{Toxic Liquidation Spirals}
\label{sec:03}
\input{sections/03_Toxic_Liquidation_Spiral}
% --------------------
\section{Mitigation Measures}
\label{sec:04}
\input{sections/04_Mitigation_Measures}
% --------------------
\section{Discussion and Conclusion}
\label{sec:05}
\input{sections/05_Conclusion}
% --------------------

% %%%%%%%%%%%%%%%%%%%%%%%%%%%%%%%%%%%%%%%%%%%%%%%%%%%%%%%%%%
% %%%%%%%%%%%%%%%%%%%%%%%%%%%%%%%%%%%%%%%%%%%%%%%%%%%%%%%%%%
% REFERENCES SECTION
% %%%%%%%%%%%%%%%%%%%%%%%%%%%%%%%%%%%%%%%%%%%%%%%%%%%%%%%%%%
% %%%%%%%%%%%%%%%%%%%%%%%%%%%%%%%%%%%%%%%%%%%%%%%%%%%%%%%%%%
%\medskip

\bibliography{references.bib} 

\newpage
\appendix
%-------------------------------
\section{Toxicity Spiral Condition}
\label{sec:06}
\input{sections/06_Toxicity_Condition}
%--------------------------------
\section{Slippage Factors}
\label{sec:07}
\input{sections/07_Slippage_Factors}
%--------------------------------
\section{Data \& Methods}
\label{sec:08}
\input{sections/08_Methods}
%--------------------------------
\end{document}

%% file: sections/01_Intro.tex
On November 13th, Avraham Eisenberg\footnote{Wallet address: 0x57E04786E231Af3343562C062E0d058F25daCE9E}, the trader linked to last month's \$114 million Mango Markets exploit, borrowed 92 million curve (CRV) tokens (worth \$38 million at the time), using 90 million USDC as collateral, on the decentralized lending platform Aave. After a series of wild swings in the CRV price, Eisenberg's position was abruptly liquidated on November 22nd. This ultimately left Aave with \$1.78 million of bad debt. In response to the attack, the Aave community is considering making changes to its liquidation threshold, implementing limitations on the number of illiquid coins that can be borrowed on the platform, and curtailing rehypothecation. Aave rehypothecates collateral posted by its clients, which increases capital efficiency but also exposes the protocol to the risk of not being able to liquidate collateral in the event of a price drop. In addition, Llama and Gauntlet authored a proposal suggesting that Aave's reserve fund and Gauntlet's insolvency fund could be used together to cover the outstanding debt. Aave's Protocol has about \$165 million in its reserve fund, while Gauntlet's has about 4,923 Aave tokens worth about \$283,000 in total. The proposal is currently under review for a governance vote.

On DeFi lending markets, users are liquidated whenever their loan-to-value (LTV) ratio surpasses a threshold value. Once that takes place, the protocol's algorithm incentivizes the repayment of the user's loans. It does so by allowing anyone to purchase the user's collateral funds at a discount. In this paper we quantify how this incentive mechanism can behave sub-optimally, and cause bad debt accrual. We further show how insolvency risks can be managed with small tweaks to the liquidation logic. We quantitatively study the statistical consequences of alternatively halting all liquidations past a certain point, adjusting the liquidation incentive dynamically as a function of the user's LTV, and modulating a technical parameter known as the closing factor. 

In Section~\ref{sec:02} we summarize the salient features of how a liquidation functions with a minimal model. In Section~\ref{sec:03} we introduce a limit beyond which the liquidation logic itself will deterministically accrue bad debt to the protocol. As a case study, we review the mechanics of the bad debt incurred by AAVE on November 22nd by employing the $\emptyset\textrm{VIX}$ protocol simulator~\cite{chaudhary2022market}. Using a mixture of all available on-chain data and minute-level price histories, all theoretical results discussed are confirmed through extensive numerical simulations. To avoid singling out the AAVE protocol, we note that the conditions for enabling such toxic liquidation spirals are actually shared ny a number of major DeFi lending markets, thus deserving significant attention. 

This work aims to extend the limited but rapidly growing literature on systematic stability in decentralized finance. Recent quantitative literature has focused on the stability of automatic market makers (AMMs)\cite{lehar2022systemic} and their value towards liquidity providers versus retaining optionality over funds~\cite{milionis2022automated}. The interplay of AMMs and lending markets is has gathered interest recently due to a number of ways it can  exhibit dynamical fragility due to a price-liquidity feedback exacerbated by informational asymmetry~\cite{chiu2022inherent}. Our research adds the role of liquidation logic and risky borrowing to understand the insolvency risk carried by DeFi lending protocols.

%% file: sections/02_Liquidation_Mechanics.tex
To liquidate a given user, liquidators are required to first repay some amount $\Delta B$ of the user's total loan $B$ with their own funds before the protocol allows them to repossess some amount $\Delta C$ of the user's collateral $C$. Generally speaking, the precise amounts paid by/to the liquidator are the result of an optimization problem whose complex details fall outside the scope of this letter. For our purposes, it will suffice the reader to know that the collateral value~\footnote{All values are expressed using the US Dollar-\$ as numeraire.} repossessed is equal to the loan amount repaid plus a premium known as the \emph{liquidation incentive} which we will represent mathematically with the letter $i$. 

For completeness, the reader should know that a liquidator is limited in how much of a user loan they are allowed to repay. This limit is imposed through a protocol-set parameter known as the \emph{closing factor}, which will be represented mathematically with the lower-case letter $c$. 

The relationship between $\Delta C$, $\Delta B$, $B$, $i$, and $c$ can be expressed mathematically as:

\bea
\Delta B &<& c\cdot B \label{eq:close_condition}\\
\Delta C &=& (1+i)\cdot\Delta B, \label{eq:incentive_condition}
\eea
where $i$ is the liquidation incentive. This is how liquidations function in a nutshell.

Liquidation incentives often vary depending on which collateral asset the liquidator wishes to repossess. More exotic assets are typically assigned larger liquidation incentives to urge liquidators to repossess a user's riskiest assets first, before focusing on safer assets such as major stablecoins (USDC, USDT) and bluechip tokens (ETH, wBTC).

In Avi's case, his portfolio only consisted of USDC collateral and CRV loans. As such, liquidators making liquidation calls to his portfolio only had the option to repay some amount of his CRV loan to repossess USDC from his collateral to capture a protocol-set liquidation incentive of 4.5\%.

The aim of liquidations is to make the portfolio of a risky user healthier. The health of a user is defined by comparing his portfolio's $LTV$ to a threshold value $LTV_{liq}$ above which the protocol will allow liquidators to intervene. Each collateral asset on a lending market has its own protocol-set threshold value, from which the user's specific $LTV_{liq}$ is computed by performing a weighted average across the user's available collateral assets.

\begin{figure}%[h]
    \centering
    \includegraphics[scale=0.5,width=\textwidth]{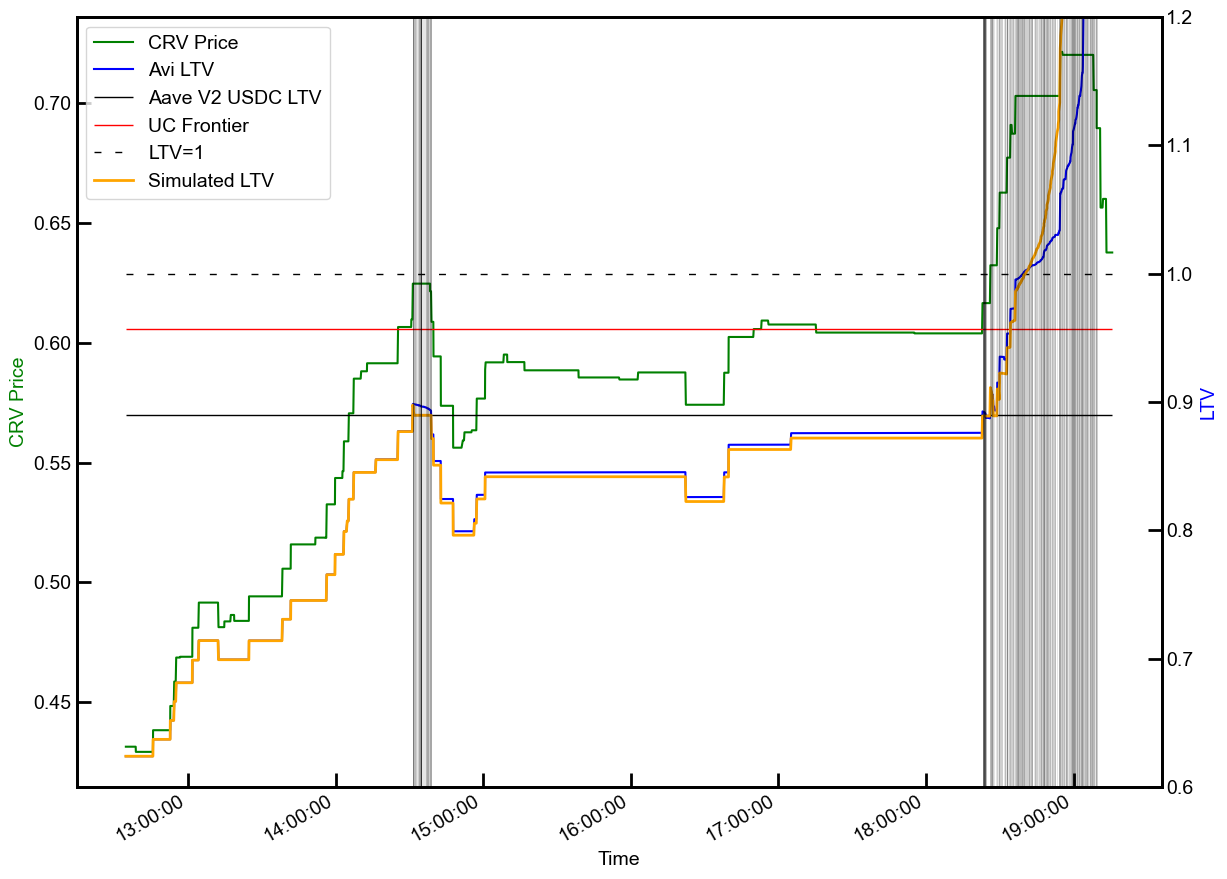}
    \caption{Avi's loan-to-value (LTV) ratio (blue; right axis) and CRV/USDC price (green; left axis) as a function of time on November 22nd, 2022. The plot follows a 6-hour timespan of activity leading to the bad debt creation event. Our simulator's reproduction of Avi's portfolio LTV is shown in gold. A thin black horizontal line marks the 89\% LTV threshold above which Avi becomes liquidatable. A red horizontal line marks the threshold beyond which liquidations become toxic (the undercollateralization frontier). A black dashed line shows $LTV = 1$ above which the user's portfolio becomes undercollateralized. Once Avi's LTV crosses the UC frontier, his LTV is worsened by each new liquidation instead of being made healthier. Upon crossing this threshold, his portfolio was guaranteed to become undercollateralized, incurring bad debt for the protocol in the process. Liquidation cascade events are shown as vertical thin lines.}
    \label{fig:1}
\end{figure}

In the case at hand, as Avi only held USDC as collateral, his portfolio's liquidation threshold was equivalent to AAVE's liquidation LTV threshold for USDC: 89\%. Whenever Avi's portfolio's loan-to-value ratio satisfied $LTV>0.89$ (the thin horizontal black line in Figure~\ref{fig:1}) liquidations would be allowed to commence (thin vertical black lines in Figure~\ref{fig:1}). The reader should appreciate that when liquidations are allowed to commence, the liquidated user's portfolio is abundantly overcollateralized (i.e. total value of the collateral is greater than the total value of loans $C>B$). This is a necessary condition for trustless lending markets to operate safely.

%% file: sections/03_Toxic_Liquidation_Spiral.tex
If at any point, the liquidation incentives lead the liquidated user's $LTV$ to worsen as a result of liquidations taking place, we will denote the liquidation as {\it toxic}. Toxic liquidations are dangerous for the protocol since they mathematically guarantee that the user's portfolio health will worsen through no fault of their own. Let us now see when this may occur.

Denote a user's initial and final loan-to-value as $LTV_{init}$ and $LTV_{fin}$ respectively. From our overview of liquidation mechanics (Equation~\ref{eq:incentive_condition} specifically), we have that:

\be
\label{eq:final_LTV}
LTV_{fin}=\frac{B-\Delta B}{C-\Delta C}=\frac{B-\Delta B}{C-(1+i)\Delta B}=\frac{B/C-(\Delta B/C)}{1-(1+i)(\Delta B/C)}=\frac{LTV_{init}-(\Delta B/C)}{1-(1+i)(\Delta B/C)},
\ee
where the values $B$, $C$, and $\Delta B$ can be considered constant once the liquidation is initiated as the entire operation takes place in one single block.

As just defined, the liquidation will be considered {\it toxic} if the final loan-to-value of the user is larger after the liquidation takes place, $LTV_{fin}>LTV_{init}$. Plugging~\ref{eq:final_LTV} into this condition~\footnote{Detailed steps can be found in Appendix~\ref{sec:06}} results in a fundamental condition between the user's initial LTV and the liquidation incentive offered to liquidators by the protocol for liquidation to be toxic:

\be
\label{eq:UC_condition}
LTV_{init}>\frac{1}{1+i}
\ee

If at {\emph ANY} point the user is liquidated while this condition holds, the user's LTV will be made worse by the liquidation. Barring some sudden and very fortuitous price action in the user's favor, this will guarantees that every successive liquidation will have the exact same effect. Liquidations will proceed until all the user's collateral has been used to repay their loans. The leftover loans once all collateral has been repossessed by liquidators will be the final bad debt incurred by the protocol. We will denote this fundamental threshold the undercollateralization (UC) frontier $LTV_{UC}$.

In Avi's case, his portfolio's constant liquidation incentive of 4.5\% implied:

\be
\label{eq:LTV_UC}
LTV_{UC} = \frac{1}{1+0.045}\simeq 0.9569 = 95.69\%.
\ee

The reader should appreciate that $LTV_{UC}<1$. This means that when the toxic liquidation spiral commences, the user's portfolio is still overcollateralized. The user has enough collateral to still cover all their loans (i.e. there is no bad debt). However, once $LTV>LTV_{UC}$, even if asset prices were to remain static, the user's portfolio will be guaranteed to become undercollateralized entirely as a result of the liquidation incentives enforced by the protocol, thus incurring bad debt.

\begin{figure}[H]
    \centering
    \includegraphics[scale=0.5,width=\textwidth]{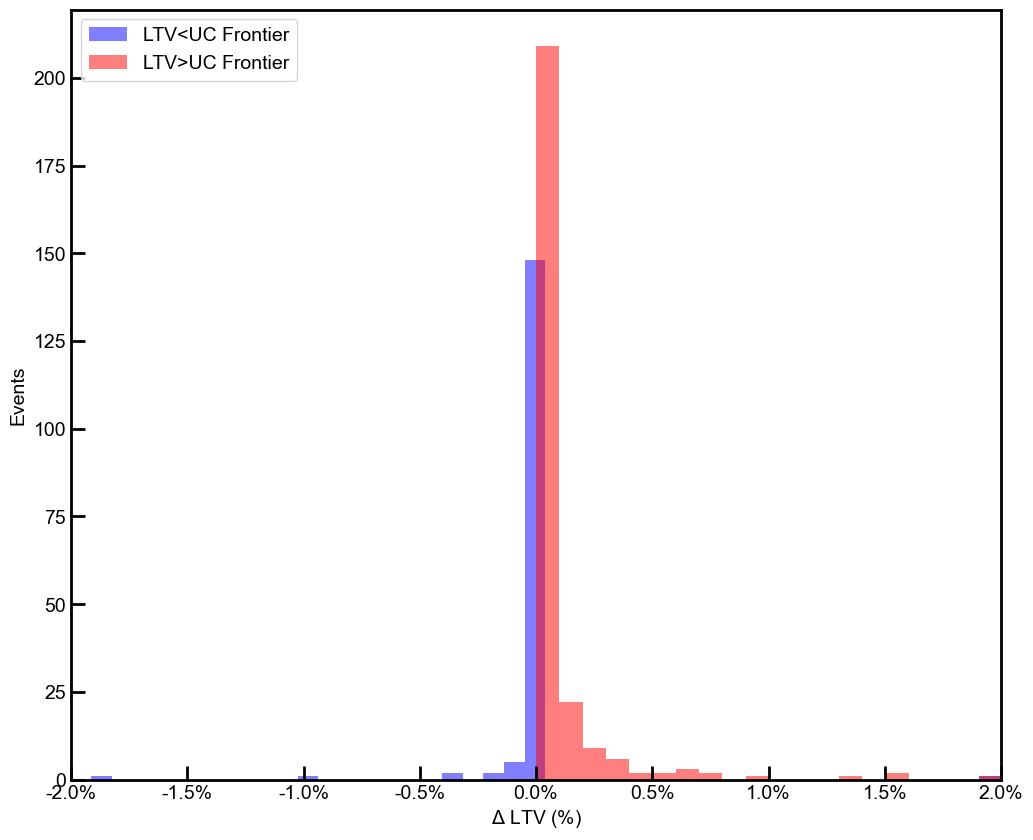}
    \caption{Distribution of changes in user's portfolio $\Delta LTV/LTV_{init} = (LTV_{fin}/LTV_{init})-1$ resulting from liquidations when $LTV<LTV_{UC}$ (blue) and when $LTV>LTV_{UC}$ (red).}
    \label{fig:2}
\end{figure}

In Figure~\ref{fig:1} the reader can see how Avi's $LTV$ evolved throughout the day. Whereas the LTV mostly changed proportionally to changes in the CRV/USDC price, once the LTV crossed the $LTV_{UC}$ threshold (horizontal red line) his LTV skyrocketed independently of the CRV/USDC price. In Figure~\ref{fig:2} one can see in detail how radically the statistics of Avi's portfolio's LTV adjustments $\Delta LTV = LTV_{fin}-LTV_{init}$ change as his loan-to-value crosses the $LTV_{UC}$ threshold.   

%% file: sections/04_Mitigation_Measures.tex
The preceding section should evoke an obvious question in the reader's mind:

\begin{center}
\emph{Should have AAVE halted liquidations?}
\end{center}

\begin{figure}[H]
    \centering
    \includegraphics[scale=0.5,width=\textwidth]{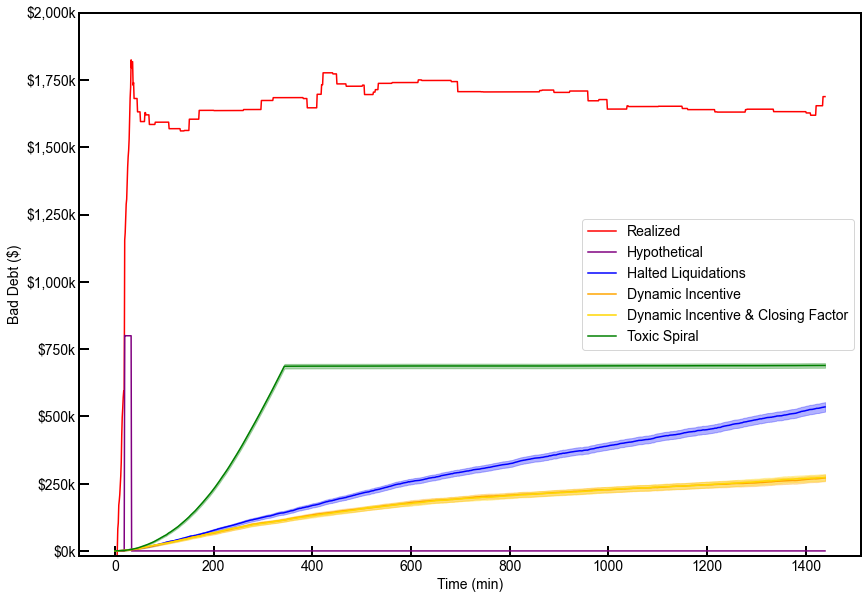}
    \caption{Bad debt incurred by AAVE as a function of time in the 24 hours (1440 minutes) following the moment where Avi's loan-to-value ratio crossed the undercollateralization threshold $LTV>LTV_{UC}$ (Equation~\ref{eq:LTV_UC}) implied by AAVE's static liquidation incentives for USDC collateral assets. \underline{\emph{Red}}: Bad debt AAVE realized due to the liquidation spirals affecting Avi Eisenberg's portfolio health. \underline{\emph{Purple}}: Bad Debt AAVE would have incurred if liquidations had been simply halted once Avi's $LTV>LTV_{UC}$. \underline{\emph{Green}}: Average bad debt Aave could have expected from the toxic liquidation spirals as per the current protocol policy. \underline{\emph{Blue}}: Average bad debt that AAVE could have statistically expected to incur if liquidations had been halted (simulation performed over $20k$ CRV/USDC historical price trajectories). \underline{\emph{Gold/Orange}}: Average bad debt that AAVE could have statistically expected to incur if liquidations had been handled using dynamic incentives alone (Orange) and with dynamic closing factor policies (Gold) (simulation performed over $20k$ CRV/USDC historical price trajectories).}
    \label{fig:3}
\end{figure}

In short, the answer is YES. By the time the toxic liquidation spiral commenced CRV/USDC prices were close to topping out after a massive 75\% run-up in prices earlier that day. Had AAVE halted all liquidations once Avi's $LTV>LTV_{UC}$, his portfolio would've momentarily become undercollateralized (to the tune of $\sim$\$750k) before immediately returning to a healthy state on its own once the CRV/USDC ultimately corrected downwards. This can be seen in the purple line on Figure~\ref{fig:3} showing what the bad debt incurred by AAVE would've been on that specific day if liquidations had simply been halted instead of allowing the toxic liquidation spiral to take place. 

Alas though, hindsight is 20/20, and nobody at the time could have possibly known how the CRV/USDC price was going to behave from that moment forward. It does however lead to a more meaningful statistical question. 

\begin{center}
\emph{How much bad debt could AAVE have expected to incur moving forward if liquidations had been halted?}
\end{center} 

At the first $LTV>LTV_{UC}$ moment there are many possible price histories that could develop. The price could in principle keep pumping forever, forcing AAVE to incur massive amounts of bad debt (significantly larger than what was actually realized). The price could also suddenly dump (as it actually happened) leading to no bad debt whatsoever. The price could also stabilize Avi's $LTV$, or make it oscillate up and down enough for healthy liquidations to take place at intervals. Both of these latter scenarios could lead to some finite amount of bad debt less than what AAVE actually incurred from the toxic liquidation spiral. Anything could've been possible, the question is: \emph{how likely would it have been?}

 The market risk assessment methodology described in the  \cite{chaudhary2022market} allows us to do precisely this. In Figure~\ref{fig:1}, the solid gold line is our liquidation simulator's reproduction of Avi's price history. The reader can appreciate how faithfully it tracks the real behavior of Avi's LTV (shown in blue). A deviation can be seen in our simulator's liquidation module after Avi's portfolio becomes undercollateralized (i.e. crosses the dashed horizontal black line). This is due to our modelling of slippages incurred by our fictitious liquidators~\footnote{More details on our slippage modelling can be found in Appendix~\ref{sec:07}.} who follow an on-chain state at each block. Overall though, our simulated reproduction of Avi's LTV is satisfactory in the overcollateralized regime, which the main thesis of this paper focuses on. In the undercollateralized regime~\footnote{When undercollateralization ($LTV>1$) is reached, outstanding loans amount to \$ 12.9M.}, our liquidation module appears to execute liquidations more efficiently than what happened in real life. As such, statistical results pertaining to toxic liquidation spirals should be deemed as optimistic.

We can use historical CRV/USDC prices to simulate alternative price histories (results are shown for 20k distinct price simulations)~\footnote{Refer to Appendix~\ref{sec:08} for more info on data and methods used}. In each, we halt/enable liquidations depending on whether, at any given moment, Avi's $LTV$ is greater/less than $LTV_{UC}$. Through each individual price trajectory simulation, we track any bad debt incurred and analyze statistics across 20k distinct runs. Readers can see the result of these simulations in the blue curve in Figure~\ref{fig:3}, where the growing shading around the curve represents 95\% confidence bands on the estimated average. At the time Avi's $LTV$ first exceeded $LTV_{UC}$, AAVE could've expected to incur $\sim$\$500k over the following 24 hours. This is roughly significantly less bad debt than what the protocol actually assured itself by allowing liquidations to proceed along their toxic spiral (red curve in Figure~\ref{fig:3}). Significantly though, the median bad debt of our simulations is ZERO (see Figure~\ref{fig:4} discussion). In the majority of simulations, the stress-testing of Avi's portfolio would not have actually incurred any bad debt whatsoever. All for something as simple as halting liquidations.

\subsection{Dynamic incentives}

The result of halting liquidations is enlightening in its simplicity, but the simulated bad debt can be improved even further with slight tweaks to the liquidation logic. Ultimately, Equation~\ref{eq:UC_condition} can be flipped on its head to obtain the largest allowable liquidation incentive given the user's $LTV$ at the moment of being liquidated, such that the liquidation is not toxic. The condition reads:

\be
\label{eq:safety_condition}
i < \frac{1}{LTV}-1
\ee

As long as the liquidation incentive satisfies this condition, user liquidations will always proceed in a healthy manner. It is important to note that when the user's portfolio becomes borderline undercollateralized ($LTV=1$) the liquidation incentive vanishes altogether. This is a safer condition than halting liquidations altogether as some liquidators may still find it profitable to liquidate a position by arbitraging the lending market's oracle's price feed. To avoid complications leading to nonsensically negative incentives when $LTV>1$, and to impose some maximal protocol-set incentive $i_0$ for a given collateral asset, the full model for healthy liquidation incentives can be written as:

\be
\label{eq:dyn_incentive}
i(LTV,i_0) = \max\left[\min\left[i_0, \frac{1}{LTV}-1-\epsilon\right],0\right]
\ee
where, for extra safety, we have introduced a static modulation parameter $\epsilon$ to guarantee that the incentive is strictly less than the right-hand side of condition~\ref{eq:safety_condition}. For practical purposes, $\epsilon$ can be any arbitrarily small, non-zero number.

\subsection{Dynamic closing factors}

The dynamic liquidation incentive~\ref{eq:dyn_incentive} can be further paired with a dynamic closing factor which increases as the user's $LTV$ inches towards unity. The idea here is that progressively larger portions of a user's portfolio should be allowed to be closed as the user's portfolio comes progressively closer to becoming undercollateralized. Whenever $LTV\geq 1$, liquidators (or protocol safety modules) should be allowed to liquidate entire asset positions of the user's portfolio in one go. There are many ways this can be expressed mathematically. For concreteness and simplicity we have tested the following linear model:

\be 
\label{eq:dyn_closing}
c(LTV,c_0) = \min\left[c_0\cdot\frac{1-LTV}{1-LTV_{liq}}+\frac{LTV-LTV_{liq}}{1-LTV_{liq}}, 1\right]
\ee
where we introduce the minimum protocol-set closing factor $c_0$ similarly to what was done for liquidation incentives earlier, as well as a $\min[\cdot,1]$ operation to guarantee that closing factors are always a number $c\leq1$. The reader can verify on their own that whenever $LTV=LTV_{liq}$ the closing factor becomes $c(LTV_{liq},c_0)=c_0$, while when $LTV\geq1$ the closing factor becomes $c(1,c_0)=1$.

The choice of expressions for~\ref{eq:dyn_incentive} and~\ref{eq:dyn_closing} is not unique. They can be modified in a number of different ways and optimized for different purposes according to protocol prerogatives. Our choice is meant solely for demonstrative purposes (where the prerogative is simplicity). 

Simultaneous use of dynamic incentives and closing factors should allow the protocol to compensate for the decreasing incentives by offering liquidators more absolute liquidity to profit from as a user's $LTV$ becomes more risky to the protocol. In Figure~\ref{fig:3}, we simulate dynamic incentives both with (orange) and without (yellow) dynamic closing factors. The reader can appreciate how dynamical incentives significantly improve on simple liquidation halting, while the additional inclusion of dynamical closing factors leads to virtually identical results. The orange/gold line (and its shaded confidence interval) shows an expectation of only $\sim$\$250k of total bad debt created 24 hours into the future. This is a significant improvement in risk management that any lending market should consider adopting. 

\begin{figure}[H]
    \centering
    \includegraphics[scale=0.5,width=\textwidth]{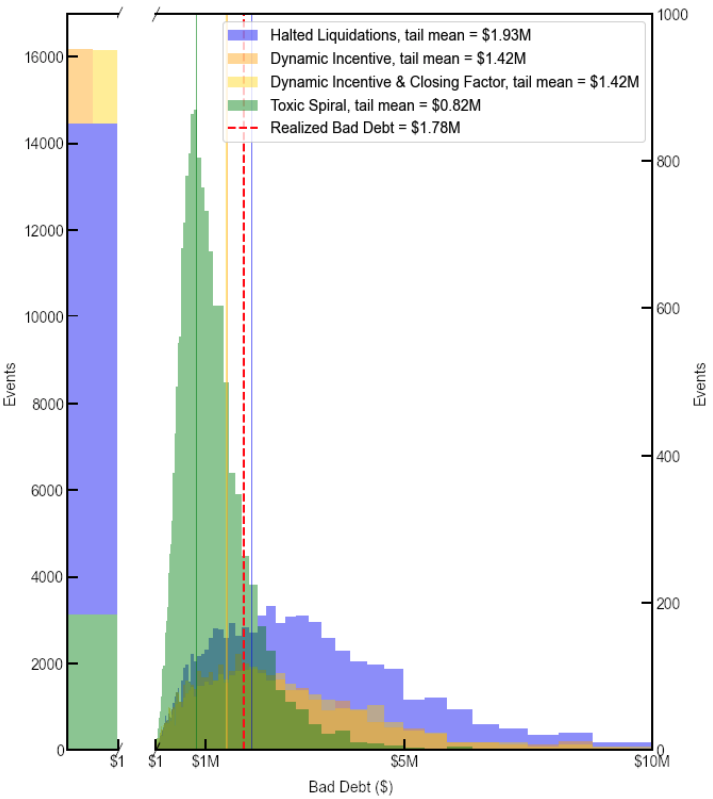}
    \caption{Distribution of bad debt after 24 hours of evolution across the various liquidation policies discussed: toxic spirals (green), halted liquidations (blue), dynamic incentives only (orange), and dynamic incentives + closing factors (gold). The x-axis is broken into two sections to highlight the magnitude of the events involving no bad debt (left vertical axis), while still offering insight into the distribution of tail events (right vertical axis).} 
    \label{fig:4}
\end{figure}

The average bad debt shown in~\ref{fig:3} does not however tell the entire story of the performance of the different policies shown. For a more nuanced insight, one must look in detail at the distribution of bad debt across the 20k price trajectories used. In Figure~\ref{fig:4} we show a detailed histogram of such bad debt distributions taken at the end of the 24hr simulation run. In Figure~\ref{fig:4}, we plot the histograms of bad debt incurred across all our policy simulations, at the end of the 24-hour simulation timeframe. Whereas the toxic liquidation spirals trade off higher chances of generating bad debt ($\sim 85\%$) for more certainty on its size (tail mean $=\$820k$), the mitigation policies discussed in the text offer significantly lower chances of generating bad debt ($\sim 19\%$) with slightly larger worst-case outcomes (tail mean$=\$1.4M$). This however assumes that no other open market interventions take place. The mitigation policies discussed in the text offer the protocol optionality on how they wish to handle toxic users on a case-by-case basis, whereas toxic liquidation spirals do not. 

%% file: sections/05_Conclusion.tex
We have demonstrated how the bad debt incurred by AAVE on November 22nd is not the result of speculative price action or irresponsible portfolio positioning. Rather, it is due to a fundamental flaw in the liquidation logic of the protocol which guaranteed that Avi Eisenberg's position would become undercollateralized with almost absolute certainty past some risky, but still overcollateralized portfolio health. We have termed this dynamic, a toxic liquidation spiral. Whereas the phenomenon was qualitatively described in the \href{https://blog.openzeppelin.com/compound-audit/}{2019 Compound audit}, to the author's knowledge a detailed study of its effects on insolvency risks was still lacking.

The theoretical insights led us to more deeply analyze alternative tweaks to lending market liquidation logic with the objective of minimizing the expectation of bad debt in the event of a sudden worsening of user's portfolio health~\footnote{The risk of generating bad debt can never be entirely extinguished.}. These tweaks were explored by stress-testing Avi Eisenberg's portfolio with thousands of alternative price trajectories using the $\emptyset\textrm{VIX}$ protocol simulator~\cite{chaudhary2022market}.

Our analysis results in a strong recommendation to all active lending markets to either halt liquidations past a certain user loan-to-value or enact dynamic liquidation incentives and closing factor policies for optimal results. Whereas very little statistical difference can be seen in the performance of dynamic liquidation incentive policy alone versus one that also adds dynamic closing factors, we argue that utilizing dynamic closing factors could offer benefits by allowing non-toxic emergency liquidations of entire user portfolios if necessary. These fall outside the scope of our simulations and would have to be studied on their own. Overall, our suggested policies show a welcome change to the risk profile taken by the protocol. A smaller amount of expected undercollateralized users created is a benefit to all entities involved. A number of major DeFi protocols could benefit from an active consideration of this analysis. 

Ultimately, the reason why liquidation LTV thresholds are set to conservative values is not just to allow buffer room for liquidators to aid in keeping lending markets healthy, but also to allow prices to evolve without the need for immediate short-term action. As a general rule of thumb, sudden short-minded responses to complex dynamical behaviors lead to outcomes worse than what the response set out to achieve. They should be avoided unless absolutely necessary.

%% file: sections/06_Toxicity_Condition.tex
Upon plugging Equation~\ref{eq:final_LTV} into the toxicity condition $LTV_{init}>LTV_{fin}$ one obtains:

\be
LTV_{fin}=\frac{LTV_{init}-(\Delta B/C)}{1-(1+i)(\Delta B/C)}>LTV_{init}, \nonumber
\ee
which, upon rearranging, gives:

\be
\label{eq:UC_reduced}
(\Delta B/C)\cdot\left[(1+i)\cdot LTV_{init}-1\right] > 0 
\ee

The condition~\ref{eq:UC_condition} is then obtained by noting that the term outside the parentheses in~\ref{eq:UC_reduced} is always greater than zero ($\Delta B/C>0$) and thus does not contribute to the condition being true or not. If $\Delta B=0$ it would simply imply that no liquidation is taking place. 

We are thus left with the condition:

\be 
(1+i)\cdot LTV_{init}-1 >0,
\ee
which upon rearranging gives Equation~\ref{eq:UC_condition} in the main text.

%% file: sections/07_Slippage_Factors.tex
Liquidators are required to first repay a loan with their own funds before repossessing collateral from the liquidated user's portfolio as discussed in Section~\ref{sec:02}. Since modelling liquidator funds is outside the scope of a first-order liquidation analysis, we assume that liquidators can flash-loan all required funds for no fees. 

A liquidator must thus compute the optimal amount $q_{repay}\equiv\Delta B$ they must flash-loan to repay the liquidated user's loan and initiate the liquidation process. All amounts are to be intended as denominated in USD\$.

In this Appendix, we walk the reader through the math of the simulator's liquidation module and how we extracted the empirical slippage factors going into our simulations (Figure~\ref{fig:5}).

\subsection{Liquidation Modelling}

The first condition on $q_{repay}$ is set by the protocol's closing factor $c$:
\be
q_{repay}< c\cdot B 
\ee

Where $B$ is the total dollar amount of outstanding user loans as described in the main text. Once $q_{repay}$ is repaid, the liquidator is allowed to repossess an amount of collateral $\Delta C = (1+i)\cdot q_{repay}$. Since a liquidator cannot repossess more collateral than the total amount $C$ which the user actually owns. This leads to a second condition on $q_{repay}$:
\be
(1+i)\cdot q_{repay} < C. 
\ee

Once the collateral has been repossessed, the liquidator will swap some amount $x$ to repay the initial flash-loan, incurring some net slippage due to swap routes and trading fees $s(x)$:
\bea
x\cdot (1-s(x)) &=& q_{repay} \label{eq:slippage_condition}\\
x &<& (1+i)\cdot q_{repay}. \nonumber
\eea

The liquidator's profit $\Pi$ is whatever is leftover from the operation:
\be
\Pi(q_{repay}) = (1+i)\cdot q_{repay}-x(q_{repay}), 
\ee
where $x(q_{repay})$ requires inverting Equation~\ref{eq:slippage_condition} first.

The final condition on $q_{repay}$ is that it be less than the amount $q_{repay}\leq q_{opt}$ which maximizes liquidator profit:
\be
\label{eq:opt_liquidity_condition}
\partial_{q}\Pi(q)\rvert_{q=q_{opt}}= 1+i - \partial_{q} x(q)\rvert_{q=q_{opt}} = 0
\ee

From which the three constraints defining $q_{repay}$ can be written together as:

\be
\label{eq:q_repay}
q_{repay} = \textrm{min}\{q_{opt}, c\cdot B, \frac{C}{1+i}\}
\ee

\subsection{Linear Slippage Model}

In the linear slippage model approximation, one has:

\be
s(x) = \gamma + \sigma \frac{x}{L},
\ee

where $gamma$ is the trading fee, $\sigma$ is the \emph{linear slippage factor}, $x$ is the amount being swapped, and $L$ the total available swap liquidity used for normalization.
\newline

Inverting equation~\ref{eq:slippage_condition} for $x$, one gets:

\be
x(q) = L\frac{1-\gamma}{2\sigma}\left[1-\sqrt{1-\frac{4\sigma q}{L\cdot (1-\gamma)^2}}\right], 
\ee
whose derivative computes to:

\be
\label{eq:dxq_linear}
\partial_{q}x(q) = \frac{1}{1-\gamma}\left[1-\frac{4\sigma q}{L\cdot (1-\gamma)^2}\right]^{-1/2}=\frac{1}{1-\gamma}\left[1-\frac{4\sigma q}{L(1-\gamma)^2}\right]^{-1/2}.
\ee

The optimal repay amount $q_{opt}$ can then be obtained by plugging~\ref{eq:dxq_linear} into~\ref{eq:opt_liquidity_condition} and solving. One gets:

\be
\label{eq:q_opt}
q_{opt} = L\cdot\frac{(1+i)^2(1-\gamma)^2-1}{4\sigma(1+i)^2}
\ee

\subsection{Empirical Slippage Factors}

Slippage factors $\sigma$ are model parameters that must be extracted from real-world data. Ideally, they require linearly approximating the real-world slippage curve as extracted from aggregators. Due to the unavailability of this historical data, we approach the slippage modelling in reverse.

Upon collecting all liquidation calls made on November 22nd, they can be classified according to whether $q_{repay}=q_{opt}$ or not. For each such liquidation, the slippage factor $\sigma$ can be obtained empirically from Equation~\ref{eq:q_opt} to give:

\be
\label{eq:empirical_sigma}
\sigma = \frac{(1+i)^2(1-\gamma)^2-1}{4(1+i)^2}\cdot \frac{L}{q_{repay}},
\ee
where we set $\gamma=0.003$ in line with typical on-chain AMM trading fees, and $L=\$ 190M$ in line with normally available liquidity.

\begin{figure}[H]
    \centering
    \includegraphics[scale=0.5,width=\textwidth]{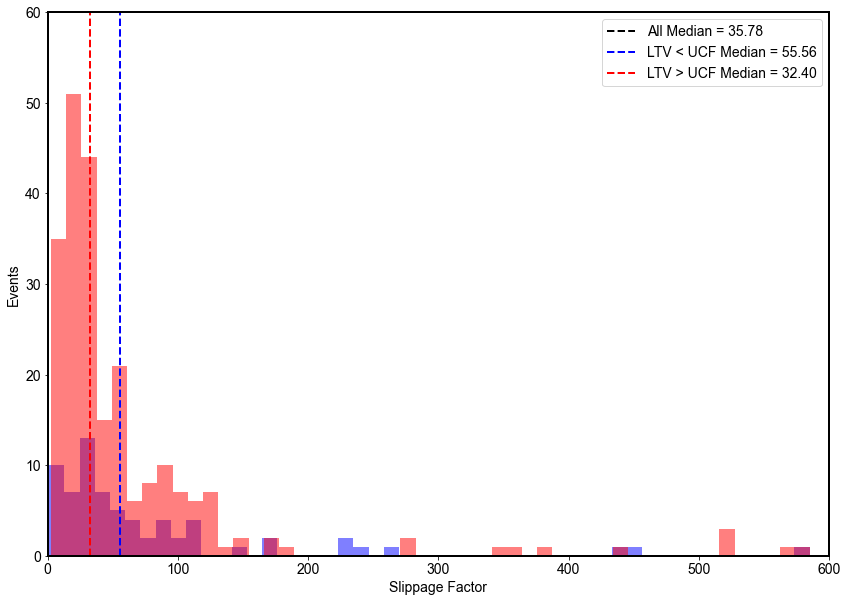}
    \caption{Independent histogram of linear slippage factors as derived from real-life liquidation events occurring before (blue) and after (red) Avi's LTV crossed the UC frontier along with their respective medians (vertical lines). The vertical black line represents the computed median of the entire data set of slippage factors, which was used as the static slippage factor in the simulator's liquidation module.}
    \label{fig:5}
\end{figure}

In Figure~\ref{fig:5} we show two histograms charting the empirical distribution of liquidation slippage factors as computed through~\ref{eq:empirical_sigma}. Following the same color-coding scheme used in Figure~\ref{fig:2}, we show the slippage factor distribution both before (blue) and after (red) Avi's portfolio crossed the UC frontier $LTV>LTV_{UC}$. Vertical dashed lines show the median slippage factor of each distribution (dashed red/blue vertical lines), as well as for the entire data as a whole (black dashed vertical line). We use the median of the entire dataset as the slippage factor in the liquidation module of the $\emptyset\textrm{VIX}$ protocol simulator~\cite{chaudhary2022market} when running simulations on Avi's portfolio's bad debt. 

For readability, Figure~\ref{fig:5} cuts the x-axis off at values of $\sigma=600$ while $22$ larger values (out of $318$ total events) were observed. 

%% file: sections/08_Methods.tex
To conduct this analysis, we queried all available on-chain data pertaining to Avi’s interactions with the AAVEv2 protocol. All historical CRV and USDC price data was collected through \href{https://www.amberdata.io/}{Amberdata}. 

The on-chain data was collected from the protocol’s \href{https://thegraph.com/hosted-service/subgraph/aave/protocol-v2?version=current}{subgraph}. We made static calls to Aave’s smart contracts at selective timestamps to confirm the reliability of \href{https://thegraph.com/}{TheGraph}’s data. Our initial dataset was comprised of 385 liquidation calls on Avi’s position, starting on November 22, 2022, 1:31:23 PM GMT, and ending on Tuesday, November 22, 2022, 6:09:23 PM. We have enriched it, also using the same source, by collecting the CRV and USDC prices at each block, beginning 2 hours before the first liquidation call and ending 24 hours after the last one. Given that the prices can fluctuate across oracles, we have decided that relying on the protocol's own will be the most reasonable approach. 

As per the asset prices, we had access to the minute-tick OHLCV data from December 11, 2021, to the day of Avi’s strategy execution. We transformed this raw data into logarithmic minute-level returns, from which 10k, 24-hour-long samples were drawn for our simulations (1440 price returns per trajectory). To overcome the negative bias trend resulting from the past 12 months of CRV price history, each price trajectory was also reversed and used to collect simulation data. This led to 20k total price trajectories used in this study.

We use an agent-based simulation of crypto money markets with Aave V2 parameters obtained by querying the protocol’s smart contracts. For a more detailed discussion, refer to the original paper on the $\emptyset\textrm{VIX}$ protocol simulator~\cite{chaudhary2022market}. 

%% file: main.bbl
\begin{thebibliography}{}

\bibitem [\protect \citeauthoryear {%
Chaudhary%
\ \BBA {} Pinna%
}{%
Chaudhary%
\ \BBA {} Pinna%
}{%
{\protect \APACyear {2022}}%
}]{%
chaudhary2022market}
\APACinsertmetastar {%
chaudhary2022market}%
\begin{APACrefauthors}%
Chaudhary, A.%
\BCBT {}\ \BBA {} Pinna, D.%
\end{APACrefauthors}%
\unskip\
\newblock
\APACrefYearMonthDay{2022}{}{}.
\newblock
{\BBOQ}\APACrefatitle {Market risk assessment: A multi-asset, agent-based
  approach applied to the DeFi lending protocols} {Market risk assessment: A
  multi-asset, agent-based approach applied to the defi lending
  protocols}.{\BBCQ}
\newblock
\APACjournalVolNumPages{\href{https://arxiv.org/abs/2211.08870}{arXiv:2211.08870}}{}{}{}.
\PrintBackRefs{\CurrentBib}

\bibitem [\protect \citeauthoryear {%
Chiu%
, Ozdenoren%
, Yuan%
\BCBL {}\ \BBA {} Zhang%
}{%
Chiu%
\ \protect \BOthers {.}}{%
{\protect \APACyear {2022}}%
}]{%
chiu2022inherent}
\APACinsertmetastar {%
chiu2022inherent}%
\begin{APACrefauthors}%
Chiu, J.%
, Ozdenoren, E.%
, Yuan, K.%
\BCBL {}\ \BBA {} Zhang, S.%
\end{APACrefauthors}%
\unskip\
\newblock
\APACrefYearMonthDay{2022}{}{}.
\newblock
{\BBOQ}\APACrefatitle {On the Inherent Fragility of DeFi Lending} {On the
  inherent fragility of defi lending}.{\BBCQ}
\newblock

\PrintBackRefs{\CurrentBib}

\bibitem [\protect \citeauthoryear {%
Lehar%
\ \BBA {} Parlour%
}{%
Lehar%
\ \BBA {} Parlour%
}{%
{\protect \APACyear {2022}}%
}]{%
lehar2022systemic}
\APACinsertmetastar {%
lehar2022systemic}%
\begin{APACrefauthors}%
Lehar, A.%
\BCBT {}\ \BBA {} Parlour, C\BPBI A.%
\end{APACrefauthors}%
\unskip\
\newblock
\APACrefYearMonthDay{2022}{}{}.
\newblock
{\BBOQ}\APACrefatitle {Systemic Fragility in Decentralized Markets} {Systemic
  fragility in decentralized markets}.{\BBCQ}
\newblock
\APACjournalVolNumPages{Available at SSRN}{}{}{}.
\PrintBackRefs{\CurrentBib}

\bibitem [\protect \citeauthoryear {%
Milionis%
, Moallemi%
, Roughgarden%
\BCBL {}\ \BBA {} Zhang%
}{%
Milionis%
\ \protect \BOthers {.}}{%
{\protect \APACyear {2022}}%
}]{%
milionis2022automated}
\APACinsertmetastar {%
milionis2022automated}%
\begin{APACrefauthors}%
Milionis, J.%
, Moallemi, C\BPBI C.%
, Roughgarden, T.%
\BCBL {}\ \BBA {} Zhang, A\BPBI L.%
\end{APACrefauthors}%
\unskip\
\newblock
\APACrefYearMonthDay{2022}{}{}.
\newblock
{\BBOQ}\APACrefatitle {Automated market making and loss-versus-rebalancing}
  {Automated market making and loss-versus-rebalancing}.{\BBCQ}
\newblock
\APACjournalVolNumPages{arXiv preprint arXiv:2208.06046}{}{}{}.
\PrintBackRefs{\CurrentBib}

\end{thebibliography}
